\documentclass[sigconf]{acmart}
\pdfoutput=1 

\AtBeginDocument{%
  \providecommand\BibTeX{{%
    \normalfont B\kern-0.5em{\scshape i\kern-0.25em b}\kern-0.8em\TeX}}}

\setcopyright{acmcopyright}
\copyrightyear{2018}
\acmYear{2018}
\acmDOI{10.1145/1122445.1122456}

\usepackage{enumitem}
\acmConference[Woodstock '18]{Woodstock '18: ACM Symposium on Neural
  Gaze Detection}{June 03--05, 2018}{Woodstock, NY}
\acmBooktitle{Woodstock '18: ACM Symposium on Neural Gaze Detection,
  June 03--05, 2018, Woodstock, NY}
\acmPrice{15.00}
\acmISBN{978-1-4503-XXXX-X/18/06}

\begin{document}

\title{Multi-Modal Summary Generation using Multi-Objective Optimization}


\author{Anubhav Jangra}
\affiliation{%
  \institution{Indian Institute of Technology Patna, India}
  }
\email{anubhav0603@gmail.com}

\author{Sriparna Saha}
\affiliation{%
  \institution{Indian Institute of Technology Patna, India}
}
\email{sriparna.saha@gmail.com}

\author{Adam Jatowt}
\affiliation{%
  \institution{Kyoto University, Japan}
}
\email{jatowt@gmail.com}

\author{Mohammad Hasanuzzaman}
\affiliation{%
  \institution{Cork Institute of Technology, Ireland}
  }
\email{hasanuzzaman.im@gmail.com}

\renewcommand{\shortauthors}{Trovato and Tobin, et al.}
\begin{abstract}
  Significant development of communication technology over the past few years has motivated research in multi-modal summarization techniques. A majority of the previous works on multi-modal summarization focus on text and images. In this paper, we propose a novel extractive multi-objective optimization based  model to produce a multi-modal summary containing text, images, and videos. 
  Important objectives such as intra-modality salience, cross-modal redundancy and cross-modal similarity are optimized simultaneously in a multi-objective optimization framework to produce effective multi-modal output. The proposed model has been evaluated separately for different modalities, and has been found to perform better than state-of-the-art approaches.
  
\end{abstract}


%
\keywords{multi-modal summarization, multi-objective optimization, differential evolution}

\maketitle

\section{Introduction}
Recent years have shown a massive outburst of multi-media content over the internet, and thus accessing/extracting useful information has become increasingly difficult.  Multi-media summarization can alleviate this problem by extracting the crux of data, and discarding the redundant or useless information. A multi-modal form of knowledge representation has several advantages over uni-modal representation of content, as it gives a more complete overview of summarized content, and provides diverse perspectives on the same topic. Having multiple forms of content representation helps reinforce ideas more concretely. Multi-modal summarization can help target a larger set of diverse reader groups, ranging from skilled readers looking to skim the information to users who are less proficient in reading and comprehending complex texts \cite{uzzaman2011multimodal}. 

\begin{figure*}
    \centering
    \includegraphics[width=\textwidth, scale=0.6]{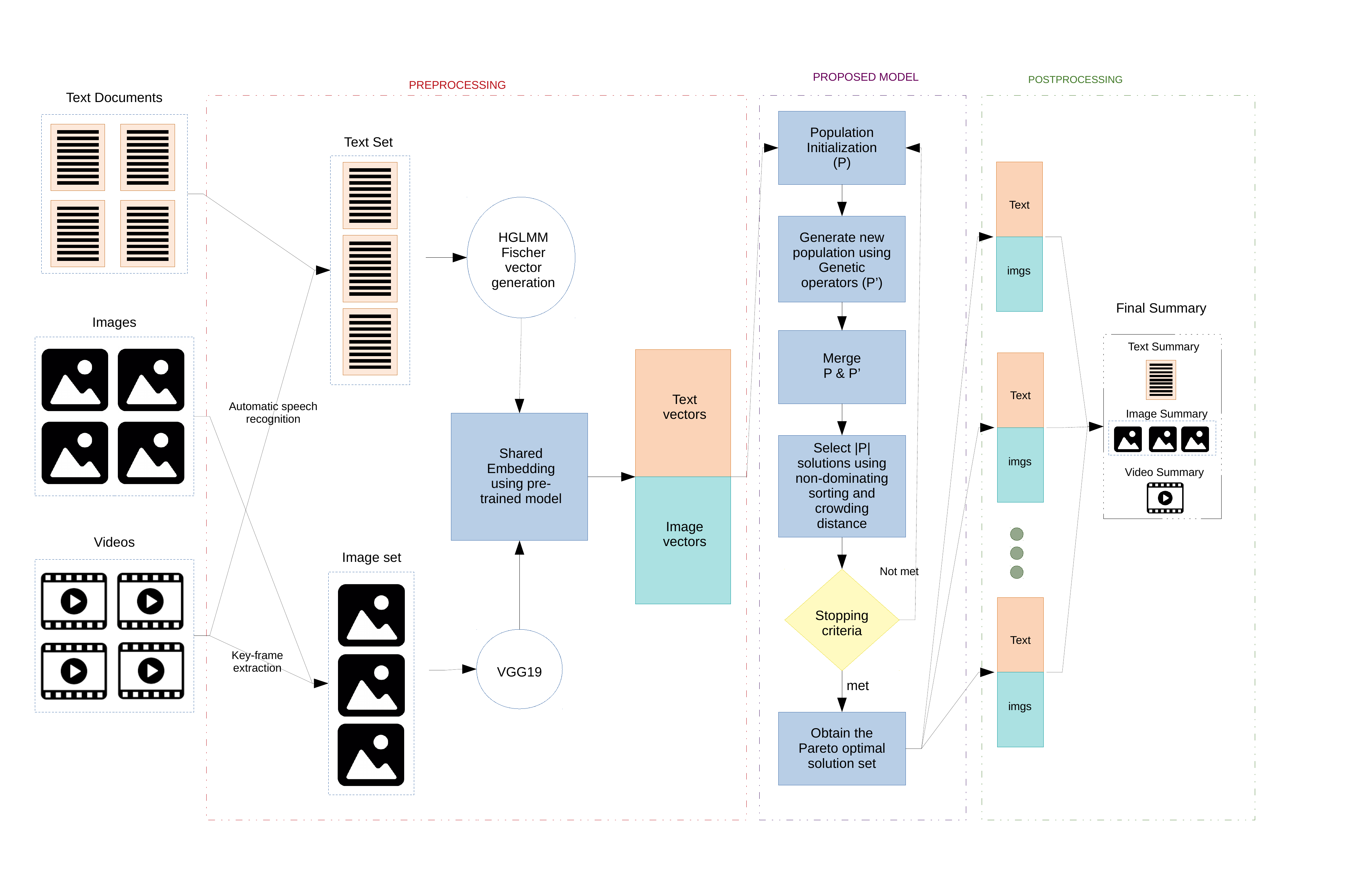}
    \caption{Proposed model architecture.}
    \label{fig:arch}
\end{figure*}

Experiments conducted by \cite{zhu2018msmo} illustrate that a multi-modal form of representation in the summary improves user satisfaction by 12.4\% compared to text summary. Most of the research work in the past has however focused on uni-modal summarization \cite{gambhir2017recent, yao2017recent}, be it text or images. Multi-modal summarization poses different challenges as one needs to also take into account the relevance between different modalities. In this paper, we focus on the task of text-image-video summary generation (TIVS) proposed by \cite{jangra2020tivs}. Unlike most text-image summarization researches, we use asynchronous data for which there is no alignment among different modalities. We propose a novel differential evolution based multi-modal summarization model using multi-objective optimization (DE-MMS-MOO). The framework of our model is shown in Fig \ref{fig:arch} 
and the main contributions of this work are as follows:
\begin{itemize}[leftmargin=3mm]
    \item This is the first attempt to solve TIVS task using multi-objective optimization (MOO) framework. MOO helps in simultaneous optimization of different objective functions like cohesion in individual modalities and consistency between multiple modalities.
    \item The proposed framework is generic. Any MOO technique can be used as the underlying optimization strategy. Here we selected a differential evolution (DE) based optimization technique, since it has been recently established that DE performs much better compared to other meta-heuristic optimization techniques \cite{das2007automatic}.
    \item The proposed model considers multimodal input (text, images, videos) and produces multimodal output (text, images and videos) with variable size of output summary. 
\end{itemize}
\vspace{-0.5em}
\section{Related Work}
Many methods have been proposed in the field of text summarization, both extractive and abstractive \cite{gambhir2017recent, yao2017recent}. Researchers have tried different approaches to tackle this problem, ranging from using integer linear programming \cite{alguliev2010multi}, deep learning models \cite{zhang2016multiview}, to graph-based techniques \cite{erkan2004lexrank} etc. 
Research on the joint representation of various modalities \cite{wang2016learning} has made the field of multi-modal information retrieval feasible. Multi-modal summarization techniques vary from abstractive text summary generation using asynchronous multi-modal data \cite{li2017multi}, to abstractive text-image summarization using deep neural networks \cite{zhu2018msmo}. Some research works have used genetic algorithms for text summarization \cite{saini2019extractive}, yet, to the best of our knowledge, no one has ever used multi-objective optimization based techniques for solving the TIVS problem. 

\vspace{-0.5em}
\section{Proposed Model}
We propose a multi-objective optimization based differential evolution technique to tackle the TIVS problem. The proposed approach takes as an input a topic with multiple documents, images and videos, and outputs an extractive textual summary, along with selected salient images and videos\footnote{For simplicity, in the current settings we output a single video only, assuming that one video is often enough.}.

\subsection{Pre-processing}
Given a topic, we have multiple related text documents, images and videos as an input. In order to extract important features from the raw data, we fetch key-frames from the videos and combine them with existing images to form the \textit{image-set}. The audio is transcribed (using IBM Watson Speech-to-Text Service\footnote{ \url{http://www.ibm.com/watson/developercloud/speech-to-text.html}}), and the resulting transcriptions together with text form the \textit{text set}. The text in \textit{text-set} is encoded using the Hybrid Gaussian-Laplacian mixture model (HGLMM) proposed in \cite{klein2014fisher}, while the images are encoded using the VGG-19 model \cite{simonyan2014very}. These model specific encodings are next fed to a two branch neural-network \cite{wang2016learning} to have 512-dimensional image and sentence vectors.

\subsection{Main model}

\subsubsection{Population Initialization}
We can see from Table \ref{tab:rouge} that the Double K-medoid algorithm performs at least as good as the multi-modal K-medoid algorithm in all the modalities (see Section \ref{model:baselines}). Thus we initialize the population ($P$) using the double K-medoid algorithm. Each solution is represented as $[$<$C_1^{txt}$, $C_2^{txt}$, ..., $C_{|K_{txt}|}^{txt}$> : <$C_1^{img}$, $C_2^{img}$, ..., $C_{|K_{img}|}^{img}$>$]$, where $C_i^{txt}$ is the $t^{th}$ text cluster center, and $|K_{txt}|$ is the maximum cluster size for text. The image part of the solution is represented similarly. The number of clusters for text and image can vary from 2 to $|K_{text}|$ for text, and from 2 to $|K_{img}|$ for images\footnote{If the number of clusters is less than the maximum value, we pad the solution.}.

\subsubsection{Generation of off-springs}\hfill\\

\textbf{Cross-over:}
For each solution in the population, we generate a mating pool by randomly selecting $H$ solutions to create new solutions. We use Eq. \ref{eq:crossover} to generate a new offspring, $y_i^{new}$. This new solution is then repaired using Eq. \ref{eq:repair}.

\begin{equation} \label{eq:crossover}
    y_i^{new}= 
    \begin{cases}
        y_i^{current} + F \times (x_i^1 - x_i^2),& \text{if rand()} \leq CR\\
        y_i^{current},              & \text{otherwise}
    \end{cases}\\
\end{equation}
where $y_i^{current}$ is the solution for which the new offspring, $y_i^{new}$, is being evaluated; $x_i^1$, $x_i^2$ are two elements from the mating pool of the solution, $y_i^{current}$; and $CR$ is the cross-over probability. 
\begin{equation} \label{eq:repair}
    y_i^{repaired}= 
    \begin{cases}
        x_i^L \times (x_i^1 - x_i^2),& \text{if } y_i^{new} \le x^L_i\\
        x_i^U,                       & \text{elseif } y_i^{new} \ge x^U_i\\
        y_i^{new}                    & \text{otherwise}
    \end{cases}\\
\end{equation}
where $x_i^L$ and $x_i^U$ are the lower and upper bounds of the population, respectively.

\textbf{Mutation:}
For each solution, we have also used three different types of mutation operations: polynomial mutation (see Eq. \ref{eq:poly_mut}), insertion mutation and deletion mutation. Polynomial mutation help in exploring  the search space and increases the convergence rate \cite{deb2008omni}. Insertion and deletion mutations also enhance exploration capabilities by increasing and decreasing the size of solution. The clusters are re-evaluated if Insertion or Deletion mutation occurs.
\begin{equation} \label{eq:poly_mut}
    y_i^{mut}= 
    \begin{cases}
         \Bigg[2r_1+(1-2r_1) \Bigg(\frac{{x^{U}_{i}} - y^{\;''}_{i}}{{x^{U}_{i}} - {x^{L}_{i}}}\Bigg)^{\eta_m+1}\Bigg]^\frac{1}{\eta_m+1}-1 ,& \text{if } r_1 \le 0.5\\
         1- \Bigg[2-2r_1+(1-2r_1) \Bigg(\frac{y^{\;''}_{i}-{x^{L}_{i}}}{{x^{U}_{i}} - {x^{L}_{i}}}\Bigg)^{\eta_m+1}\Bigg]^\frac{1}{\eta_m + 1},              & \text{otherwise}
    \end{cases}\\
\end{equation}

\subsubsection{Selection of top $|P|$ solutions}
We use the concept of non-dominating sorting and crowding distance to select the $|P|$ best solutions \cite{deb2002fast}.

\subsubsection{Stopping criteria}
The process is repeated until the maximum number of generations ($g_{max}$) is reached.

\subsubsection{Objective functions}
We propose two different sets of objective functions, based on which we design two different models. \\
\textbf{Summarization-based:} Three objectives <Sal(txt) / Red(txt), Sal(img) / Red(img), CrossCorr> are simultaneously maximized. Salience, redundancy and cross-modal correspondence are calculated using Eq. \ref{eq:sal}, \ref{eq:red}, \ref{eq:cross}, respectively.
\\
\begin{equation}\label{eq:sal}
    Sal(mod) = \sum_{c_j^{mod}} \sum_{x_k^{mod}\in cluster^{mod}(i)} sim_{cos}(c_j^{mod}, x_k^{mod})
\end{equation}
\begin{equation}\label{eq:red}
    Red(mod) = \sum_{c_j^{mod}} \sum_{c_i^{mod}\, i\neq j} sim_{cos}(c_j^{mod}, c_i^{mod})
\end{equation}
where $mod\in\{txt, img\}$, $c_j^{mod}$ is the $j^{th}$ cluster for modality $mod$, and $cluster^{mod}$ returns the elements of $j^{th}$ cluster. 
\\
\begin{equation}\label{eq:cross}
    CrossCorr = \sum_{c_j^{txt}} \sum_{c_i^{img}\, i\neq j} sim_{cos}(c_j^{img}, c_i^{txt})
\end{equation}

\textbf{Clustering-based:} We use PBM index \cite{pakhira2004validity}, which is a popular cluster validity index (function of cluster compactness and separation) to evaluate the uni-modal clustering for text and images. Thus, we maximize three  objectives <PBM(txt), PBM(img), CrossCorr> where cross-modal correspondence is evaluated by Eq. \ref{eq:cross}.
\vspace{-0.5em}
\subsection{Post-processing}
After the termination criteria is met, the model outputs $P$ solutions containing text-image pairs of variable length. We select the Pareto optimal solutions from the population, and for each solution, we generate the text summary from the text part of the solution. In order to generate the image summary, we select those image vectors that are not key-frames, and also select those images from the initial images that have a minimum cosine similarity of $\alpha$ and maximum similarity of $\beta$ \cite{jangra2020tivs}. For comparison the values of $\alpha \text{ and }\beta$ are kept the same as in \cite{jangra2020tivs}. For each video, a weighted sum of visual and verbal scores is computed as described in \cite{jangra2020tivs}. 

\begin{table*}
  \scriptsize
  \caption{Evaluation scores for the text/image/video summary. The `-' denotes unavailability of summary for that modality due to model constraints.}
  \vspace{-3mm}
  \label{tab:rouge}
  \begin{tabular}{ccccccc}
    \toprule
    Model&ROUGE R-1&ROUGE R-2&ROUGE R-L&Image Average Precision&Image Average Recall&Video Accuracy\\
    \midrule
    Random video selection (10 attempts) &-&-&-&-&-& 16\%\\
    TextRank & 0.312 & 0.117 & 0.273 &-&-&-\\
    Image match & \textbf{0.422} & 0.133 &-&-&-&-\\ 
    JILP-MMS & 0.260 & 0.074 & 0.226& 0.599 & 0.383 & 44\%\\
    Double K-medoid & 0.248 & 0.067 & 0.213& 0.7188 & 0.9392 & 40\%\\
    multi-modal K-medoid & 0.242 & 0.061 & 0.209& 0.6137 & 0.8005 & 40\%\\
    Uni-modal optimization based DE-MMS-MOO & 0.352 & 0.183 & 0.318 & 0.7382 & 0.9435 & \textbf{44\%}\\
    Summarization-based DE-MMS-MOO & 0.405 & \textbf{0.194} & 0.370 & 0.7513 & 0.9766 & \textbf{44\%}\\
    Clustering-based DE-MMS-MOO & 0.420 & 0.167 & \textbf{0.390} & \textbf{0.7678} & \textbf{0.9822} & \textbf{44\%}\\
  \bottomrule
\end{tabular}
\end{table*}

\vspace{-1em}
\section{Experimental Setting}
\subsection{Dataset}
We use the multi-modal summarization dataset prepared by \cite{jangra2020tivs}. The dataset consists of 25 topics describing different events in the news domain. Each topic contains 20 text documents, 3-9 images and 3-8 videos. For each topic there are also three text references, and at least one image as well as one video are provided as the multi-modal summary.

\subsection{Baselines} \label{model:baselines}
We evaluate our proposed model with several strong baselines ranging from existing state-of-the-art techniques to novel approaches that we propose.

\textbf{Baseline-1 (TextRank):}
We evaluate the quality of our text summary against the graph-based TextRank algorithm \cite{mihalcea2004textrank}, by feeding it the entire text\footnote{We use Python's open-source Gensim library's implementation:  \url{https://radimrehurek.com/gensim/summarization/summariser.html}.}.

\textbf{Baseline-2 (Image match MMS):} A greedy technique to generate textual summary using multi-modal guidance strategy is proposed in \cite{li2017multi}\footnote{The paper does not report ROUGE R-L scores.}. Out of multiple variations proposed in that research, the image match seems to be the most promising, and thus we use it to compare with our model.

\textbf{Baseline-3 (JILP-MMS):}
Jangra et. al. \cite{jangra2020tivs} proposed a joint-integer linear programming based method to generate multi-modal summary. JILP-MMS model uses intra-modal salience, intra-modal diversity and inter-modal correspondence as objective functions.

\textbf{Baseline-4 (Double K-medoid):}
After the preprocessing step, we perform two separate K-medoid clustering algorithms, one for sentences and the other for images. Since the text and images share the representation space, the other modality participates in the clustering process in the sense that it cannot become the cluster center, but it can still participate in the membership calculation of each cluster. The rest of the process is the same as the standard K-medoid algorithm\footnote{For all the k-medoid steps performed in our research we applied \textit{kmeans++} seeding \cite{arthur2006k} over randomly initialized  cluster centers.}.

\textbf{Baseline-5 (multi-modal K-medoid):} \label{model:multi_modal_k_medoid}
In this method, we assume that there is one single modality, and we run the K-medoid algorithm until convergence. The top-K sentences and top-K images are selected as the data points which are nearest to the cluster centers for each of the $k$ clusters, respectively, for each modality.

\textbf{Baseline-6 (Uni-modal optimization based DE-MMS-MOO):}
We use the framework proposed in Section 3, but instead of tri-objective optimization, we instead optimize two objectives <Sal(txt) / Red(txt), Sal(img) / Red(img)>, where salience and redundancy are calculated  using Eq. \ref{eq:sal} and Eq. \ref{eq:red}, respectively. For fair comparison all the hyperparameters and model settings are kept the same for this baseline as ones for the proposed models.

\begin{figure}
    \centering
    \includegraphics[width=\linewidth, scale=0.5]{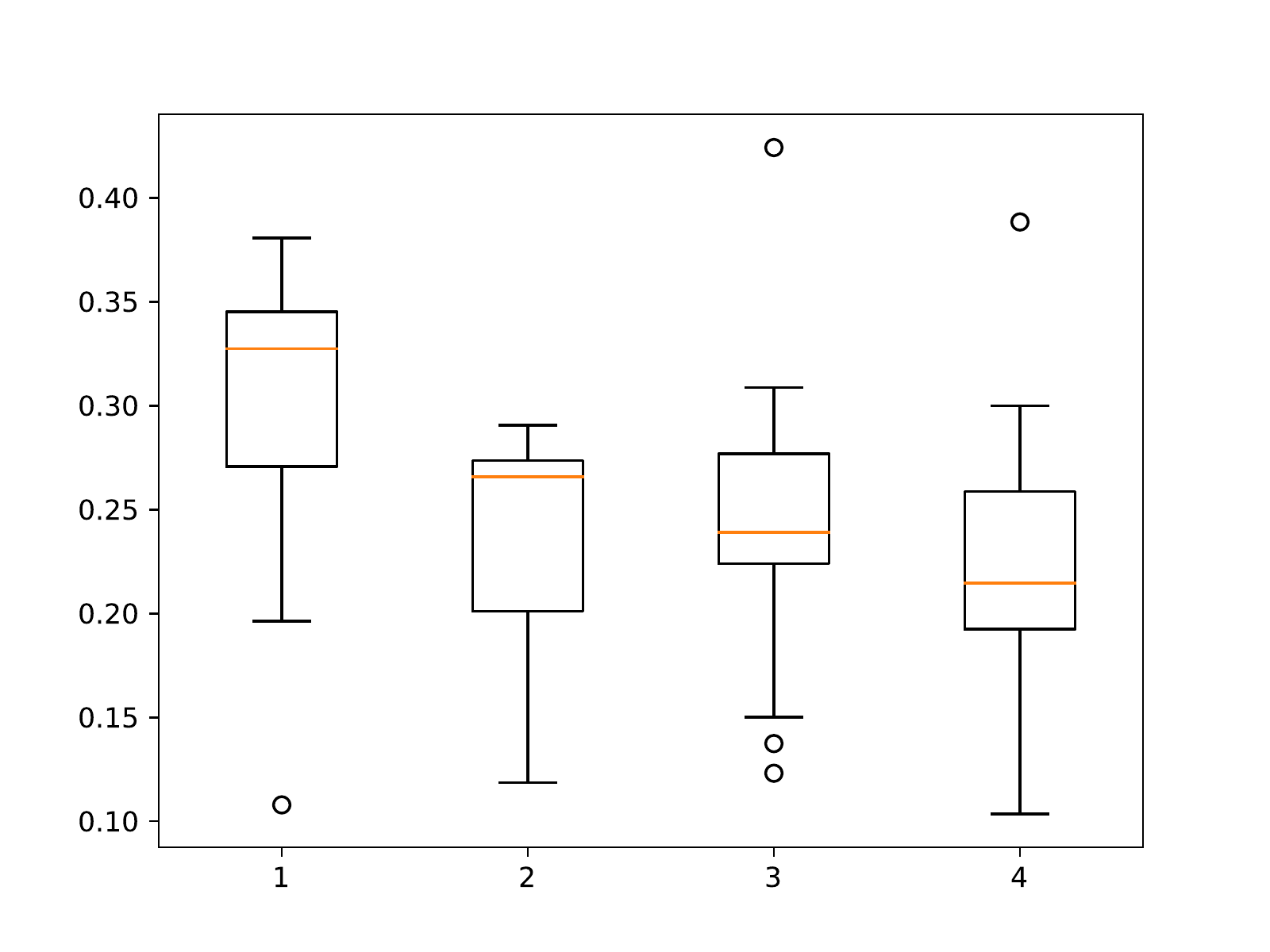}
    \vspace{-4mm}
    \caption{Box-whiskers plot for ROUGE-L values of all solutions on Pareto front for randomly selected four topics.}
    \label{fig:box}
\end{figure}

\section{Results}
Table \ref{tab:rouge} shows that our model performs better than the rest of the techniques. In order to evaluate the scores for our population based techniques, the maximum of all the evaluation scores are taken per topic, in order to compare the best of our model's ability with other baselines. The two proposed models, namely \textit{Summarization-based DE-MMS-MOO} and \textit{Clustering-based DE-MMS-MOO}, perform better than the other models in different modalities. As \textit{Double K-medoid} baseline performs better than the \textit{multi-modal K-medoid} baseline, we use this technique for solution initialization in all the proposed models. The \textit{Uni-modal optimization based DE-MMS-MOO} model works better than the clustering based baselines (\textit{Double K-medoid}, \textit{multi-modal K-medoid}). This reassures the fact that differential evolution brings about a positive change. Both the proposed models perform at least as good as the \textit{Uni-modal optimization based DE-MMS-MOO} baseline when trained under the same settings, and thus we can state that in order to generate a supplementary multi-modal summary, cross-modal correspondence is an important objective. This shows us that multiple modalities assist each other to bring out more useful information from the data. Since our model produces multiple summaries, it is important to ensue that all of the produced summaries are of good quality. To demonstrate this we draw a box-whiskers plot for ROUGE R-L score values of all the solutions on final Pareto front, for four randomly selected topics. Since all the modalities are equally significant, we cannot however directly comment on the superiority of one model over the other.


\section{Conclusion}
In this paper, we propose a novel multi-model summary generation technique that surpasses the existing state-of-the-art multi-modal summarization models. We use the proposed framework in two different objective settings, both of which have comparable performance in all the modality evaluations. Although we only explore the framework's potential using differential evolution based approaches, the proposed framework is generic and is adaptable to different settings. 

{\bf Acknowledgement}: Dr. Sriparna Saha gratefully acknowledges the Young Faculty Research Fellowship (YFRF) Award, supported by Visvesvaraya PhD scheme for Electronics and IT, Ministry of Electronics and Information Technology (MeitY), Government of India, being implemented by Digital India Corporation (formerly Media Lab Asia) for carrying out this research.   
\bibliographystyle{ACM-Reference-Format}
\bibliography{sample-base}

\end{document}